\documentclass[12pt]{article}
\usepackage{amssymb}
\usepackage{lscape}

\renewcommand{\theequation}{\arabic{section}.\arabic{equation}}

\begin{document}

%************************** Text Begins here ******************************

%  Greek letters

\def\a{\alpha}
\def\b{\beta}
\def\d{\delta}
\def\e{\epsilon}
\def\g{\gamma}
\def\h{\mathfrak{h}}
\def\k{\kappa}
\def\l{\lambda}
\def\o{\omega}
\def\p{\wp}
\def\r{\rho}
\def\t{\tau}
\def\s{\sigma}
\def\z{\zeta}
\def\x{\xi}
\def\V={{{\bf\rm{V}}}}
 \def\A{{\cal{A}}}
 \def\B{{\cal{B}}}
 \def\C{{\cal{C}}}
 \def\D{{\cal{D}}}
\def\G{\Gamma}
\def\K{{\cal{K}}}
\def\O{\Omega}
\def\R{\bar{R}}
\def\T{{\cal{T}}}
\def\L{\Lambda}
\def\f{E_{\tau,\eta}(sl_2)}
\def\E{E_{\tau,\eta}(sl_n)}
\def\Zb{\mathbb{Z}}
\def\Cb{\mathbb{C}}

\def\R{\overline{R}}
% Shorthands for \begin{equation} and the like

\def\beq{\begin{equation}}
\def\eeq{\end{equation}}
\def\bea{\begin{eqnarray}}
\def\eea{\end{eqnarray}}
\def\ba{\begin{array}}
\def\ea{\end{array}}
\def\no{\nonumber}
\def\le{\langle}
\def\re{\rangle}
\def\lt{\left}
\def\rt{\right}

\newtheorem{Theorem}{Theorem}
\newtheorem{Definition}{Definition}
\newtheorem{Proposition}{Proposition}
\newtheorem{Lemma}{Lemma}
\newtheorem{Corollary}{Corollary}
\newcommand{\proof}[1]{{\bf Proof. }
        #1\begin{flushright}$\Box$\end{flushright}}

\baselineskip=20pt

%%%%%%%%%%%%%%%%%%%%%%%%%%%%%%%%%%%%%%%%%%%%%%%%%%%%%%%%%%%%
%                                                          %
%  Title page                                              %
%                                                          %
%%%%%%%%%%%%%%%%%%%%%%%%%%%%%%%%%%%%%%%%%%%%%%%%%%%%%%%%%%%%
\newfont{\elevenmib}{cmmib10 scaled\magstep1}
\newcommand{\preprint}{
   \begin{flushleft}
     %\elevenmib Yukawa\, Institute\, Kyoto\\
   \end{flushleft}\vspace{-1.3cm}
   \begin{flushright}\normalsize
  % \sf  YITP-03-53\\
   %  {\tt hep-th/yymmnnn} \\ November 2005
   \end{flushright}}
\newcommand{\Title}[1]{{\baselineskip=26pt
   \begin{center} \Large \bf #1 \\ \ \\ \end{center}}}
\newcommand{\Author}{\begin{center}
   \large \bf
Junpeng Cao${}^{a,b}$,~Shuai Cui${}^{a}$,~Wen-Li Yang${}^{c,d}\footnote{Corresponding author:
wlyang@nwu.edu.cn}$,~Kangjie Shi${}^c$~ and~Yupeng
Wang${}^{a,b}\footnote{Corresponding author: yupeng@iphy.ac.cn}$
 \end{center}}
\newcommand{\Address}{\begin{center}

     ${}^a$Beijing National Laboratory for Condensed Matter
           Physics, Institute of Physics, Chinese Academy of Sciences, Beijing
           100190, China\\
     ${}^b$Collaborative Innovation Center of Quantum Matter, Beijing,
     China\\
     ${}^c$Institute of Modern Physics, Northwest University,
     Xian 710069, China\\
     ${}^d$Beijing Center for Mathematics and Information Interdisciplinary Sciences, Beijing, 100048,  China

   \end{center}}
\newcommand{\Accepted}[1]{\begin{center}
   {\large \sf #1}\\ \vspace{1mm}{\small \sf Accepted for Publication}
   \end{center}}

\preprint
\thispagestyle{empty}
\bigskip\bigskip\bigskip

\Title{Spin-${\frac 12}$ XYZ model revisit: general solutions via off-diagonal Bethe ansatz } \Author

\Address
\vspace{1cm}

\begin{abstract}
The spin-${\frac 12}$ XYZ model with both periodic and anti-periodic
boundary conditions is studied via the off-diagonal Bethe ansatz
method. The exact spectra of the Hamiltonians and the Bethe ansatz equations are derived by
constructing the inhomogeneous $T-Q$ relations, which allow us to
treat both the even $N$ (the number of lattice sites) and odd $N$
cases simultaneously in an unified approach.

\vspace{1truecm} \noindent {\it PACS:}
75.10.Pq, 03.65.Vf, 71.10.Pm

\end{abstract}
\newpage

%%%%%%%%%%%%%%%%%%%%%%%%%%%%%%%%%%%%%%%%%%%%%%%%%%%%%%%%%%%%%%
%                                                             %
%  1. Introduction                                            %
%                                                             %
%%%%%%%%%%%%%%%%%%%%%%%%%%%%%%%%%%%%%%%%%%%%%%%%%%%%%%%%%%%%%%%
\section{Introduction}
\label{intro} \setcounter{equation}{0}

The spin-$\frac 12$ XYZ model is a typical model in statistical
physics, one-dimensional magnetism and quantum communication.
The first exact solution of the model with periodic boundary
condition was derived by Baxter
\cite{2-baxter1-1,2-baxter1-2,2-baxter1-3,2-baxter1-4} based on its
intrinsic relationship with the classical two-dimensional eight-vertex
model. In his famous series works, the fundamental equation (the
Yang-Baxter equation \cite{yang1,yang2,bax2}) was emphasized and the
$T-Q$ method was proposed. Subsequently, Takhtadzhan and Faddeev
\cite{xyz2} resolved the model by the algebraic Bethe ansatz method
\cite{sk,yb}. In both Baxter's and Takhatadzhan and Faddeev's
approaches, local gauge transformation played a very  important role
in obtaining a proper local vacuum state (or reference state) with
which the general Bethe states can be constructed.
However, a proper reference state is so far only available for even
$N$ (the number of lattice sites) but not for odd $N$. This constitutes the main obstacle
for applying the conventional Bethe ansatz methods to the latter
case. In fact, the lack of a reference state is a common feature of
the integrable models without $U(1)$ symmetry and has been a very
important and difficult issue in the field of quantum integrable
models.
%To deal with
%such kind of models, a promising method is the off-diagonal Bethe
%ansatz \cite{cao1}, with which an extended T-Q relation can be
%constructed based on the operator product identities without using
%the information of states.

In this paper, we revisit the XYZ model by employing the
off-diagonal Bethe ansatz (ODBA) method proposed recently by the present
authors \cite{cao1,cao2,cao2-1}.
The  Hamiltonian of the XYZ spin chain is
\begin{eqnarray}
H = \frac{1}{2}\sum_{n=1}^N
  (J_x\sigma_n^x\sigma_{n+1}^x+J_y\sigma_n^y\sigma_{n+1}^y
  +J_z\sigma_n^z\sigma_{n+1}^z). \label{xyzh}
\end{eqnarray}
The coupling constants are parameterized as
\begin{eqnarray}
J_x =e^{i\pi\eta}\frac{\sigma(\eta+\frac{\tau}{2})}{\sigma(\frac{\tau}{2})}, \quad
J_y =e^{i\pi\eta}\frac{\sigma(\eta+\frac{1+\tau}{2})}{\sigma(\frac{1+\tau}{2})},\quad
J_z= \frac{\sigma(\eta+\frac{1}{2})}{\sigma(\frac{1}{2})},\label{Coupling}
\end{eqnarray}
with the elliptic function $\sigma(u)$ defined by (\ref{Function})
below and $\sigma^x,\,\sigma^y,\,\sigma^z$ being the usual Pauli
matrices. The Hamiltonian with either periodic boundary condition
\bea \sigma^{x}_{N+1}=\sigma^{x}_{1},\quad
\sigma^{y}_{N+1}=\sigma^{y}_{1},\quad
\sigma^{z}_{N+1}=\sigma^{z}_{1},\label{Periodic-BC} \eea or
anti-periodic boundary condition (or the quantum topological spin
ring \cite{cao1}) \bea
  \sigma^x_{N+1}=\sigma^x_1,\quad   \sigma^y_{N+1}=-\sigma^y_1,\quad
  \sigma^z_{N+1}=-\sigma^z_1,
  \label{Anti-periodic}
\eea is integrable.

The paper is organized as follows.  Section 2 serves as an
introduction of our notations and some basic ingredients. After
briefly reviewing  the inhomogeneous XYZ spin chain with periodic
boundary condition, we derive the operator product identities of the
transfer matrix  at some special points of the spectral parameter.
In Section 3, the inhomogeneous $T-Q$ relation for the eigenvalues
of the transfer matrix and the corresponding Bethe ansatz equations
(BAEs) are constructed based on the operator product identities  of
the transfer matrix and its  quasi-periodic properties. Section 4 is
attributed to the exact solution of the XYZ spin chain with
antiperiodic boundary condition. In Section 5, we summarize our
results. Some useful identities about the elliptic functions are
listed in Appendix A. The trigonometric limit is given in Appendix
B.

%%%%%%%%%%%%%%%%%%%%%%%%%%%%%%%%%%%%%%%%%%%%%%%%%%%%%%%%%%%%%%%
%                                                             %
%  2. Transfer matrix                                         %
%                                                             %
%                                                             %
%                                                             %
%%%%%%%%%%%%%%%%%%%%%%%%%%%%%%%%%%%%%%%%%%%%%%%%%%%%%%%%%%%%%%%

\section{ Transfer matrix}
\label{XYZ} \setcounter{equation}{0}

Let us fix a generic complex number $\eta$ and a generic complex number $\tau$ such that ${\rm
Im}(\tau)>0$. For convenience,
we introduce the following elliptic functions
\begin{eqnarray} &&\theta\left[
\begin{array}{c}
a_1\\a_2
\end{array}\right](u,\tau)=\sum_{m=-\infty}^{\infty}
\exp\left\{i\pi\left[(m+a_1)^2\tau+2(m+a_1)(u+a_2)\right]\right\},\label{Function-a-b}\\[6pt]
&&\s(u)=\theta\left[\begin{array}{c}\frac{1}{2}\\[2pt]\frac{1}{2}
\end{array}\right](u,\tau),\quad \zeta(u)=\frac{\partial}{\partial u}
\left\{\ln\s(u)\right\}.\label{Function}
\end{eqnarray}
The well-known $R$-matrix for the eight-vertex model,  $R(u)\in {\rm
End}(\Cb^2\otimes \Cb^2)$ is given by \bea
R(u)=\lt(\begin{array}{llll}\alpha(u)&&&\delta(u)\\&\beta(u)&\gamma(u)&\\
&\gamma(u)&\beta(u)&\\ \delta(u)&&&\alpha(u)\end{array}\rt),
\label{r-matrix}\eea with the non-zero entries \cite{bax2}
\bea
&&\hspace{-2.0truecm}\alpha(u)\hspace{-0.1truecm}=
 \hspace{-0.1truecm}\frac{\theta\lt[\begin{array}{c} 0\\\frac{1}{2}
 \end{array}\rt]\hspace{-0.16truecm}(u,2\tau)\hspace{0.12truecm}
 \theta\lt[\begin{array}{c} \frac{1}{2}\\[2pt]\frac{1}{2}
 \end{array}\rt]\hspace{-0.16truecm}(u+\eta,2\tau)}{\theta\lt[\begin{array}{c} 0\\\frac{1}{2}
 \end{array}\rt]\hspace{-0.16truecm}(0,2\tau)\hspace{0.12truecm}
 \theta\lt[\begin{array}{c} \frac{1}{2}\\[2pt]\frac{1}{2}
 \end{array}\rt]\hspace{-0.16truecm}(\eta,2\tau)},\quad
\beta(u)\hspace{-0.1truecm}=\hspace{-0.1truecm}\frac{\theta\lt[\begin{array}{c}
 \frac{1}{2}\\[2pt]\frac{1}{2}
 \end{array}\rt]\hspace{-0.16truecm}(u,2\tau)\hspace{0.12truecm}
 \theta\lt[\begin{array}{c} 0\\\frac{1}{2}
 \end{array}\rt]\hspace{-0.16truecm}(u+\eta,2\tau)}
 {\theta\lt[\begin{array}{c} 0\\\frac{1}{2}
 \end{array}\rt]\hspace{-0.16truecm}(0,2\tau)\hspace{0.12truecm}
 \theta\lt[\begin{array}{c} \frac{1}{2}\\[2pt]\frac{1}{2}
 \end{array}\rt]\hspace{-0.16truecm}(\eta,2\tau)},\no\\[6pt]
&&\hspace{-2.0truecm}\gamma(u)\hspace{-0.1truecm}=
 \hspace{-0.1truecm}\frac{\theta\lt[\begin{array}{c} 0\\\frac{1}{2}
 \end{array}\rt]\hspace{-0.16truecm}(u,2\tau)\hspace{0.12truecm}
 \theta\lt[\begin{array}{c} 0\\\frac{1}{2}
 \end{array}\rt]\hspace{-0.16truecm}(u+\eta,2\tau)}
 {\theta\lt[\begin{array}{c} 0\\\frac{1}{2}
 \end{array}\rt]\hspace{-0.16truecm}(0,2\tau)\hspace{0.12truecm}
 \theta\lt[\begin{array}{c} 0\\\frac{1}{2}
 \end{array}\rt]\hspace{-0.16truecm}(\eta,2\tau)},\quad
\delta(u)\hspace{-0.1truecm}=\hspace{-0.1truecm}\frac{\theta\lt[\begin{array}{c}
 \frac{1}{2}\\[2pt]\frac{1}{2}
 \end{array}\rt]\hspace{-0.16truecm}(u,2\tau)\hspace{0.12truecm}
 \theta\lt[\begin{array}{c} \frac{1}{2}\\[2pt]\frac{1}{2}
 \end{array}\rt]\hspace{-0.16truecm}(u+\eta,2\tau)}
 {\theta\lt[\begin{array}{c} 0\\\frac{1}{2}
 \end{array}\rt]\hspace{-0.16truecm}(0,2\tau)\hspace{0.12truecm}
 \theta\lt[\begin{array}{c} 0\\\frac{1}{2}
 \end{array}\rt]\hspace{-0.16truecm}(\eta,2\tau)}.\label{r-func}
\eea

\noindent Here $u$ is the spectral parameter and $\eta$ is the
crossing parameter. In addition to  satisfying the quantum
Yang-Baxter equation (QYBE),
\begin{eqnarray}
\hspace{-1.2truecm}R_{12}(u_1-u_2)R_{13}(u_1-u_3)R_{23}(u_2-u_3)=R_{23}(u_2-u_3)R_{13}(u_1-u_3)R_{12}(u_1-u_2),\label{QYB}
\end{eqnarray}
the $R$-matrix also possesses the following properties \bea
&&\hspace{-1.5cm}\mbox{ Initial
condition}:\,R_{12}(0)= P_{12},\label{Int-R}\\[6pt]
&&\hspace{-1.5cm}\mbox{ Unitarity
relation}:\,R_{12}(u)R_{21}(-u)= -\xi(u)\,{\rm id},
\quad \xi(u)=\frac{\s(u-\eta)\s(u+\eta)}{\s(\eta)\s(\eta)},\label{Unitarity}\\[6pt]
&&\hspace{-1.5cm}\mbox{ Crossing
relation}:\,R_{12}(u)=V_1R_{12}^{t_2}(-u-\eta)V_1,\quad
V=-i\s^y,
\label{crosing-unitarity}\\[6pt]
&&\hspace{-1.5cm}\mbox{ PT-symmetry}:\,R_{12}(u)=R_{21}(u)=R^{t_1\,t_2}_{12}(u),\label{PT}\\[6pt]
&&\hspace{-1.5cm}\mbox{$Z_2$-symmetry}: \,
\qquad\s^i_1\s^i_2R_{1,2}(u)=R_{1,2}(u)\s^i_1\s^i_2,\quad
\mbox{for}\,\,
i=x,y,z,\label{Z2-sym}\\[6pt]
&&\hspace{-1.5cm}\mbox{Antisymmetry}:\,R_{12}(-\eta)=-(1-P_{12})=-2P_{12}^{(-)}.\label{Ant}
\eea Here $R_{21}(u)=P_{12}R_{12}(u)P_{12}$ with $P_{12}$ being the
usual permutation operator and $t_i$ denotes transposition in the
$i$-th space. Throughout this paper we adopt the standard notations:
for any matrix $A\in {\rm End}(\Cb^2)$, $A_j$ is an embedding
operator in the tensor space $\Cb^2\otimes \Cb^2\otimes\cdots$,
which acts as $A$ on the $j$-th space and as identity on the other
factor spaces; $R_{i\,j}(u)$ is an embedding operator of $R$-matrix
in the tensor space, which acts as identity on the factor spaces
except for the $i$-th and $j$-th ones.

Let us introduce the monodromy matrix
\begin{eqnarray}
T_0(u)&=&R_{0N}(u-\theta_N)\ldots R_{01}(u-\theta_1),\label{Mon-1}
\end{eqnarray}
where $\{\theta_j|j=1,\cdots,N\}$ are generic  free complex
parameters which are usually called inhomogeneous parameters. The
transfer matrix $t(u)$ of the inhomogeneous  XYZ chain with periodic boundary
condition (\ref{Periodic-BC}) is given by \cite{bax2}
 \begin{eqnarray}
 t(u)=tr_0\lt\{T_0(u)\rt\},\label{trans-per}
 \end{eqnarray} where $tr_0$ denotes the trace over the
``auxiliary space" $0$. The Hamiltonian (\ref{xyzh}) with the periodic boundary condition is given by
\begin{eqnarray}
H=\frac{\sigma(\eta)}{\sigma'(0)}\left\{\frac{\partial \ln
t(u)}{\partial
u}|_{u=0,\theta_j=0}-\frac{1}{2}N\zeta(\eta)\right\},\label{ham}
\end{eqnarray}
where $\sigma'(0)=\left.\frac{\partial}{\partial
u}\,\sigma(u)\right|_{u=0}$ and the function $\zeta(u)$ is given by
(\ref{Function}). It is remarked that the identities
(\ref{Identity-fuction})-(\ref{Identity-f-4}) (see Appendix A) are
very useful to give the expressions (\ref{Coupling}). The QYBE
(\ref{QYB}) leads to that the transfer matrices with different
spectral parameters are mutually commutative \cite{yb}, i.e.,
$[t(u),t(v)]=0$, which guarantees the integrability of the model by
treating $t(u)$ as the generating functional of the conserved
quantities.

Let us evaluate the transfer matrix of the closed chain at some
special points. The initial condition of the $R$-matrix
(\ref{Int-R}) implies that
\begin{eqnarray}
t(\theta_j)&=&R_{j\,j-1}(\theta_j-\theta_{j-1})\ldots R_{j\,1}(\theta_j-\theta_{1})\nonumber\\[6pt]
&&\times R_{j\,N}(\theta_j-\theta_{N})\ldots
R_{j\,j+1}(\theta_j-\theta_{j+1}).\label{T1}
\end{eqnarray} The crossing relation (\ref{crosing-unitarity}) enables one to have
\begin{eqnarray}
t(\theta_j-\eta) &=&(-1)^NR_{j\,j+1}(-\theta_j
+\theta_{j+1})\ldots R_{j\,N}
(-\theta_j+\theta_{N})\nonumber\\[6pt]
&&\times R_{j\,1}(-\theta_j+\theta_{1})\ldots
R_{j\,j-1}(-\theta_j+\theta_{j-1}).\label{T22}
\end{eqnarray} With (\ref{T1})-(\ref{T22}) and the unitary relation (\ref{Unitarity}) we readily obtain
the following operator identities
\begin{eqnarray}
t(\theta_j)t(\theta_j-\eta)=\Delta_q(\theta_j),\,j=1,\ldots,N,\label{Main-identity-1}
\end{eqnarray}
where the quantum determinant $\Delta(u)$ of the monodromy matrix is
proportional to the identity operator
\begin{eqnarray}
&&\Delta_q(u)=a(u)d(u-\eta)\times {\rm id},\\[6pt]
&&a(u)=\prod_{l=1}^N
\frac{\sigma(u-\theta_l+\eta)} {\sigma(\eta)},\quad d(u)=a(u-\eta)=
\prod_{l=1}^N
\frac{\sigma(u-\theta_l)} {\sigma(\eta)}.\label{a-d-functions}
\end{eqnarray} In addition, (\ref{Unitarity}) and (\ref{T1}) give rise to the following
operator identity \cite{Kit99, Mai00,Goh00} \bea
\prod_{j=1}^N\,t(\theta_j)=\prod_{j=1}^N a(\theta_j)\times {\rm
id}.\label{Main-identity-2} \eea The $Z_2$-symmetry (\ref{Z2-sym})
of the $R$-matrix implies \bea
&&U^i\,t(u)\,U^i=tr_0\lt(U^i\,T_0(u)\,U^i\rt)=
tr_0\lt(\s^i_0\,T_0(u)\,\s^i_0\rt)=t(u),
  \label{Invariant-1}\\[6pt]
&& U^i=\s^i_1\s^i_2\ldots \s^i_N,\quad i=x,y,z.\label{Invarinat-2}
\eea Notice that $\{U^i\}$ form an (non)abelian group when $N$ is
even (odd), i.e. \bea (U^i)^2={\rm id},\quad
U^i\,U^j=(-1)^NU^j\,U^i,\,\, {\rm for}\,\, i\neq j, \,\,{\rm
and}\,\,i,j=x,y,z. \label{Invarinat-3} \eea

The quasi-periodicity of the $\s$-function \bea
\sigma(u+\tau)=-e^{-2i\pi(u+\frac{\tau}{2})}\sigma(u),\quad
\sigma(u+1)=-\sigma(u),\label{Quasi-function} \eea indicates that
the $R$-matrix possesses the following quasi-periodic properties\bea
R_{12}(u+1)&=&-\s^z_1R_{12}(u)\s^z_1,\no\\[6pt]
R_{12}(u+\tau)&=&-e^{-2i\pi(u+\frac{\eta}{2}+\frac{\tau}{2})}
\s^x_1R_{12}(u)\s^x_1,\no \eea which lead to the quasi-periodicity
of the transfer matrix $t(u)$
\begin{eqnarray}
&&t(u+\tau)=(-1)^Ne^{-2\pi
i\{Nu+N(\frac{\eta+\tau}{2})-\sum_{j=1}^N\theta_j\}} t(u),\label{quasi-transfer-1}\\[6pt]
&&t(u+1)=(-1)^Nt(u).\label{quasi-transfer-2}
\end{eqnarray}
In the subsequent section we shall show that (\ref{Main-identity-1}), (\ref{Main-identity-2}) and
(\ref{quasi-transfer-1})-(\ref{quasi-transfer-2}), allow us to
determine the eigenvalue $\Lambda(u)$ of the transfer matrix $t(u)$
completely.

%%%%%%%%%%%%%%%%%%%%%%%%%%%%%%%%%%%%%%%%%%%%%%%%%%%%%%%%%%%%%%%
%                                                             %
%  3. Functional relations and the T-Q relation               %
%                                                             %
%                                                             %
%                                                             %
%%%%%%%%%%%%%%%%%%%%%%%%%%%%%%%%%%%%%%%%%%%%%%%%%%%%%%%%%%%%%%%

\section{Functional relations and the $T-Q$ relation}
\label{BAE}
\setcounter{equation}{0}

Let $|\Psi\rangle$ be an eigenstate (independent of $u$) of $t(u)$
with the eigenvalue $\Lambda(u)$, i.e., \bea
t(u)|\Psi\rangle=\Lambda(u)|\Psi\rangle.\no \eea  The analyticity of
the $R$-matrix  implies that
\begin{eqnarray}
 \Lambda(u) \mbox{ is an entire
function of $u$}.\label{Eigen-Per-1}
\end{eqnarray}
The quasi-periodic properties of the transfer matrix
(\ref{quasi-transfer-1}) and (\ref{quasi-transfer-2}) indicate that
the corresponding eigenvalue $\Lambda(u)$ also possesses the
following quasi-periodic properties \bea
&&\Lambda(u+1)=(-1)^N\Lambda(u),\label{Eigen-Per-2}\\
&&\Lambda(u+\tau)=(-1)^Ne^{-2\pi
i\{Nu+N(\frac{\eta+\tau}{2})-\sum_{j=1}^N\theta_j\}}
\Lambda(u).\label{Eigen-Per-3} \eea The analytic property
(\ref{Eigen-Per-1}) and the quasi-periodic properties
(\ref{Eigen-Per-2})-(\ref{Eigen-Per-3}) indicate that  $\Lambda(u)$,
as a function of $u$, is an elliptic polynomial of degree $N$. This implies that one
needs $N+1$ conditions to fix the function. The very operator identities (\ref{Main-identity-1}) and
(\ref{Main-identity-2}) lead to that the corresponding eigenvalue
$\Lambda(u)$ satisfies the following relations (the same functional relations to (\ref {Eigen-identity})
were previously derived in \cite{G7} via separation of variables
method)
\begin{eqnarray}
&&\Lambda(\theta_j)\Lambda(\theta_j-\eta) =a(\theta_j)
d(\theta_j-\eta),\,j=1,\ldots,N,\label{Eigen-identity}\\[6pt]
&&\prod_{j=1}^N\Lambda(\theta_j)=\prod_{j=1}^Na(\theta_j).\label{Eigen-identity-0}
\end{eqnarray} Therefore, the equations (\ref{Eigen-Per-1})-(\ref{Eigen-identity-0}) will completely characterize the
spectrum of the transfer matrix. Following the work  \cite{cao1,cao2,cao2-1}, we can construct the following
inhomogeneous $T-Q$ relation for the eigenvalue $\Lambda(u)$
\begin{eqnarray}
\Lambda(u)&=&e^{2i\pi l_1u+i\phi}a(u)\frac{Q_1(u-\eta)Q(u-\eta)}{Q_2(u)Q(u)}
+e^{-2i\pi l_1(u+\eta)-i\phi}d(u)\frac{Q_2(u+\eta)Q(u+\eta)}{Q_1(u)Q(u)}\nonumber\\[6pt]
&&+c\,\frac{\sigma^m(u+\frac\eta2)a(u)d(u)}{\sigma^m(\eta)Q_1(u)Q_2(u)Q(u)},\label{T-Q}
\end{eqnarray}
where $l_1$ is a  certain integer and $m$ is a non-negative integer.
The functions $Q_1(u)$, $Q_2(u)$ and $Q(u)$ are parameterized by
$2M+M_1$ unequal Bethe roots $\{\mu_j|j=1,\ldots,M\}$,
$\{\nu_j|j=1,\ldots,M\}$ and $\{\l_j|j=1,\ldots,M_1\}$ as follows
\begin{eqnarray}
Q_1(u)&=&\prod_{j=1}^{M}\frac{\sigma(u-\mu_j)}{\sigma(\eta)},\,\,
Q_2(u)=\prod_{j=1}^{M}\frac{\sigma(u-\nu_j)}{\sigma(\eta)},\label{Q-1}\\[6pt]
Q(u)&=&\prod_{j=1}^{M_1}\frac{\sigma(u-\l_j)}{\sigma(\eta)}.\label{Q-2}
\end{eqnarray} These non-negative integers $m$, $M$ and $M_1$ satisfy the following relation
\bea N+m=2M+M_1.\label{Integer-Constraint} \eea It should be
remarked that the minimal number of the Bethe roots is $N$ when
$m=0$. In the following text, we put $m=0$. It is believed that any
choice of $m$ might give a complete set of eigenvalues $\L(u)$ of
the transfer matrix.

In order that the function (\ref{T-Q}) becomes the solution of
(\ref{Eigen-Per-1}) - (\ref{Eigen-Per-3}), the $N+2$ parameters
$\phi$, $c$,  $\{\mu_j\}$, $\{\nu_j\}$ and $\{\l_j\}$ have to
satisfy the following $N+2$ equations
\begin{eqnarray}
&&(\frac{N}2-M-M_1)\eta-\sum_{j=1}^M(\mu_j-\nu_j)=l_1\tau+m_1, \quad l_1,\,m_1\in Z, \label{BAE-1}\\[6pt]
&&\frac{N}{2}\eta
-\sum_{l=1}^N\theta_l+\sum_{j=1}^M(\mu_j+\nu_j)+\sum_{j=1}^{M_1}\l_j=m_2,\quad m_2\in Z,\label{BAE-2}\\[6pt]
&&c\,e^{2i\pi(l_1\mu_j+l_1\eta)+i\phi}a(\mu_j)=-Q_2(\mu_j)Q_2(\mu_j+\eta)Q(\mu_j+\eta),\quad j=1,\ldots,M,\label{BAE-3}\\[6pt]
&&c\,e^{-2i\pi l_1\nu_j-i\phi}d(\nu_j)=-Q_1(\nu_j)Q_1(\nu_j-\eta)Q(\nu_j-\eta),\quad j=1,\ldots,M,\label{BAE-4}\\[6pt]
&&\frac{e^{2i\pi l_1(2\l_j+\eta)+2i\phi}a(\l_j)}{d(\l_j)}+\frac{Q_2(\l_j)Q_2(\l_j+\eta)Q(\l_j+\eta)}
{Q_1(\l_j)Q_1(\l_j-\eta)Q(\l_j-\eta)}\no\\[6pt]
&&\qquad\qquad =\frac{-c\,e^{2i\pi l_1(\l_j+\eta)+i\phi}a(\l_j)}{Q_1(\l_j)Q_1(\l_j-\eta)Q(\l_j-\eta)},\quad j=1,\ldots,M_1.\label{BAE-5}
\end{eqnarray}
The equations (\ref{BAE-3})-(\ref{BAE-5}) ensure that the function
(\ref{T-Q}) is an entire function of $u$, namely, the function
satisfies (\ref{Eigen-Per-1}). The equations (\ref{BAE-1}) and
(\ref{BAE-2}) imply that the function (\ref{T-Q}) has the same
quasi-periodic properties to
(\ref{Eigen-Per-2})-(\ref{Eigen-Per-3}). As $\s(0)=0$, $\L(u)$ given
by (\ref{T-Q}) at the points $u=\theta_j$ and $u=\theta_j-\eta$
takes the values \bea \Lambda(\theta_j)&=&e^{2i\pi
l_1\theta_j+i\phi}a(\theta_j)\frac{Q_1(\theta_j-\eta)Q(\theta_j-\eta)}{Q_2(\theta_j)Q(\theta_j)},
\quad j=1,\ldots, N, \label{Eigen-theta}\\[6pt]
\Lambda(\theta_j-\eta)&=&e^{-2i\pi
l_1\theta_j-i\phi}d(\theta_j-\eta)\frac{Q_2(\theta_j)Q(\theta_j)}{Q_1(\theta_j-\eta)Q(\theta_j-\eta)},
\quad j=1,\ldots, N,\no \eea which directly yield
\bea
\Lambda(\theta_j)\Lambda(\theta_j-\eta) =a(\theta_j)
d(\theta_j-\eta),\quad j=1,\ldots,N,\no \eea namely, $\L(u)$ given by
(\ref{T-Q}) indeed satisfies (\ref{Eigen-Per-1}) -
(\ref{Eigen-identity}) and is the eigenvalue of the transfer matrix,
provided that the BAEs (\ref{BAE-1})-(\ref{BAE-5}) hold. Taking the
homogeneous limit $\theta_j\rightarrow 0$, the $T-Q$ relation
becomes
\begin{eqnarray}
\Lambda(u)&=&e^{2i\pi l_1u+i\phi}\frac{\sigma^N(u+\eta)}{\sigma^N(\eta)}\frac{Q_1(u-\eta)Q(u-\eta)}{Q_2(u)Q(u)}\no\\[6pt]
&&+\frac{e^{-2i\pi l_1(u+\eta)-i\phi}\sigma^N(u)}{\sigma^N(\eta)}\frac{Q_2(u+\eta)Q(u+\eta)}{Q_1(u)Q(u)}\nonumber\\[6pt]
&&+\frac{c\,\sigma^N(u+\eta)\sigma^N(u)}{Q_1(u)Q_2(u)Q(u)\sigma^N(\eta)\sigma^N(\eta)},\label{T-Q-Main}
\end{eqnarray}
with the corresponding BAEs
\begin{eqnarray}
&&\hspace{-1.2truecm}(\frac{N}2-M-M_1)\eta-\sum_{j=1}^M(\mu_j-\nu_j)=l_1\tau+m_1, \quad l_1,\, m_1\in Z, \label{BAE-Main-1}\\[6pt]
&&\hspace{-1.2truecm}\frac{N}{2}\eta
+\sum_{j=1}^M(\mu_j+\nu_j)+\sum_{j=1}^{M_1}\l_j=m_2,\quad m_2\in Z,\label{BAE-Main-2}\\[6pt]
&&\hspace{-1.2truecm}\frac{c\,e^{2i\pi(l_1\mu_j+l_1\eta)+i\phi}\sigma^N(\mu_j+\eta)}{\sigma^N(\eta)}=
-Q_2(\mu_j)Q_2(\mu_j+\eta)Q(\mu_j+\eta),\quad j=1,\ldots,M,\label{BAE-Main-3}\\[6pt]
&&\hspace{-1.2truecm}\frac{c\,e^{-2i\pi l_1\nu_j-i\phi}\sigma^N(\nu_j)}{\sigma^N(\eta)}=
-Q_1(\nu_j)Q_1(\nu_j-\eta)Q(\nu_j-\eta),\quad j=1,\ldots,M,\label{BAE-Main-4}\\[6pt]
&&\hspace{-1.2truecm}\frac{e^{2i\pi l_1(2\l_j+\eta)+2i\phi}\s^N(\l_j+\eta)}{\s^N(\l_j)}+\frac{Q_2(\l_j)Q_2(\l_j+\eta)Q(\l_j+\eta)}
{Q_1(\l_j)Q_1(\l_j-\eta)Q(\l_j-\eta)}\no\\[6pt]
&&\qquad\qquad =\frac{-c\,e^{2i\pi
l_1(\l_j+\eta)+i\phi}\s^N(\l_j+\eta)}
{Q_1(\l_j)Q_1(\l_j-\eta)Q(\l_j-\eta)\s^N(\eta)},\quad
j=1,\ldots,M_1,\label{BAE-Main-5}
\end{eqnarray}
and the selection rule \bea
\Lambda(0)=e^{i\phi}\lt\{\prod_{j=1}^M\frac{\sigma(\mu_j+\eta)}{\sigma(\nu_j)}\rt\}
\lt\{\prod_{j=1}^{M_1}\frac{\sigma(\l_j+\eta)}{\sigma(\l_j)}\rt\}
=e^{\frac{2i\pi k}{N}},\quad k=1,\ldots,N.\label{Selection-1} \eea
The eigenvalue of the Hamiltonian (\ref{xyzh}) with periodic
boundary condition is given by
\begin{eqnarray}
E=\frac{\sigma(\eta)}{\sigma'(0)}\lt\{\sum_{j=1}^M[\zeta(\nu_j)-\zeta(\mu_j+\eta)]+
\sum_{j=1}^{M_1}[\zeta(\l_j)-\zeta(\l_j+\eta)]+\frac{1}{2}N\zeta(\eta)+2i\pi
l_1\rt\}.\label{e}
\end{eqnarray}

Some remarks are in order. The integers $l_1$, $m_1$ and $m_2$ that
appeared in the BAEs (\ref{BAE-Main-1})-(\ref{BAE-Main-5}) are due to
the quasi-periodicity of the $R$-matrix
(\ref{r-matrix})-(\ref{r-func}) in terms of $u$. Any choices of these
integers may give rise to the complete set of  eigenvalues $\L(u)$.
In addition, the numerical simulation for the open XXZ chain
\cite{G9} indicates that the BAEs with a fixed $M$ (or $M_1$) indeed
give the complete solutions of the model (see also
\cite{cao3}) . Similarly, in our case, different $M$
might only give different parameterizations of the eigenvalues but
not different eigenstates. To support this conjecture, numerical
simulations for $N=3,5,4$ with random choice of $\eta$ and $\tau$
are performed. The results are listed in Table 1, 2, 3 respectively.
Moreover, (\ref{Invariant-1}) and (\ref{Invarinat-3}) imply that
$\L(u)$ has no degeneracy for even $N$ but indeed has a double
degeneracy for odd $N$. As a consequence, for the even $N$ case
there exists a one-to-one correspondence between the solutions of
the BAEs (\ref{Generic-BAE-1})-(\ref{Generic-BAE-2}) (see below) and
the eigenstates of the transfer matrix, while for the odd $N$ case
there are multiple solutions of the BAEs
(\ref{BAE-Main-1})-(\ref{Selection-1}) corresponding to one
$\L(u)$ due to its degeneracy. This phenomenon has been checked
numerically for some small $N$.

\begin{table}
\caption{Numerical solutions of the BAEs
(\ref{BAE-Main-1})-(\ref{Selection-1}) for $N=3$, $M=1$,
$\eta=0.20$, $\tau=i$, $l_1=m_1=m_2=0$. The eigenvalues $E_n$
calculated from (\ref{e}) are exactly the same to those from the
exact diagonalization of the Hamiltonian. $n$ denotes the number of the energy levels }
\begin{center}{\tiny
\begin{tabular}{ccc|c|c|c|c|c} \hline\hline
$\mu_1$ & $\nu_1$ & $\lambda_1$ & $c$ & $\phi$ & $k$ & $E_n$ & $n$\\
\hline
$0.35000+0.02632i$ & $0.45000+0.02632i$ & $-1.10000-0.05263i$ & $-0.08948+0.00000i$ & $-0.08501-0.00000i$ & $1$ & $-1.40865$ & $1$ \\
$0.35000-0.02632i$ & $0.45000-0.02632i$ & $-1.10000+0.05263i$ & $-0.08948+0.00000i$ & $0.08501-0.00000i$ & $2$ & $-1.40865$ & $1$ \\
$-0.15000+0.08693i$ & $-0.05000+0.08693i$ & $-0.10000-0.17387i$ & $3.04065+0.00000i$ & $4.10893-0.00000i$ & $2$ & $-1.40865$ & $1$ \\
$-0.15000-0.08693i$ & $-0.05000-0.08693i$ & $-0.10000+0.17387i$ & $3.04065-0.00000i$ & $-4.10893-0.00000i$ & $1$ & $-1.40865$ & $1$ \\
$-0.65000-0.27875i$ & $-0.55000-0.27875i$ & $0.90000+0.55749i$ & $-0.28951-0.00000i$ & $0.35925-0.00000i$ & $0$ & $1.18468$ & $2$ \\
$-0.28066+0.31196i$ & $-0.18066+0.31196i$ & $0.16133-0.62392i$ & $-0.61188+0.36729i$ & $-0.27657+0.04967i$ & $0$ & $1.18468$ & $2$ \\
$0.15828+0.12139i$ & $0.25828+0.12139i$ & $-0.71655-0.24279i$ & $-0.09303-0.16695i$ & $-0.29190+0.31832i$ & $0$ & $1.63263$ & $3$ \\
$-0.42198+0.50000i$ & $-0.32198+0.50000i$ & $0.443 97-1.00000i$ & $3.33371-7.57925i$ & $-0.94248-0.14392i$ & $0$ & $1.63263$ & $3$ \\
\hline\hline \end{tabular}} \end{center}\end{table}

\begin{landscape}
\begin{table}
\caption{Numerical solutions of the BAEs
(\ref{BAE-Main-1})-(\ref{Selection-1}) for $N=5$, $\eta=0.20$,
$M=1$, $\tau=i$, $l_1=m_1=m_2=0$. The eigenvalues $E_n$ calculated
from (\ref{e}) are exactly the same to those from the exact
diagonalization of the Hamiltonian. $n$ denotes the number of the energy levels. }
\begin{center}{\scriptsize
\begin{tabular}{c|c|ccc|c|c|c|c|c} \hline\hline
$\mu_1$ & $\nu_1$ & $\l_1$ & $\l_2$ & $\l_3$ & $c$ &
$\phi$ & $k$ & $E_n$ & $n$\\ \hline
$-0.55827-0.02265i$ & $-0.25827-0.02265i$ & $-0.10018+0.08190i$ & $-0.10011-0.01038i$ & $0.51684-0.02622i$ & $-0.08617-0.00699i$ & $0.10696-0.03689i$ & $1$ & $-3.51343$ & $1$ \\
$-0.55827+0.02265i$ & $-0.25827+0.02265i$ & $-0.10018-0.08190i$ & $-0.10011+0.01038i$ & $0.51684+0.02622i$ & $-0.08617+0.00699i$ & $-0.10696-0.03689i$ & $4$ & $-3.51343$ & $1$ \\
$-0.35211+0.07575i$ & $-0.05211+0.07575i$ & $-2.07865+0.00704i$ & $0.90992-0.07849i$ & $1.07296-0.08006i$ & $-0.26973+6.70848i$ & $1.26914+1.97113i$ & $1$ & $-3.51343$ & $1$ \\
$-0.35211-0.07575i$ & $-0.05211-0.07575i$ & $-2.07865-0.00704i$ & $0.90992+0.07849i$ & $1.07296+0.08006i$ & $-0.26973-6.70848i$ & $-1.26914+1.97113i$ & $4$ & $-3.51343$ & $1$ \\
$-1.42021-0.00000i$ & $-1.12021-0.00000i$ & $-0.06738+0.09747i$ & $2.17518+0.00000i$ & $-0.06738-0.09747i$ & $-3.95699+0.00000i$ & $0.00000+1.74063i$ & $0$ & $-1.42192$ & $2$ \\
$0.44197-0.00000i$ & $0.74197-0.00000i$ & $-4.09984+0.09712i$ & $-1.09984-0.09712i$ & $3.51573-0.00000i$ & $-0.09625-0.00000i$ & $-0.00000-0.02711i$ & $0$ & $-1.42192$ & $2$ \\
$-0.37881-0.02198i$ & $-0.07881-0.02198i$ & $-0.08454+0.09207i$ & $-0.02749-0.11816i$ & $0.06967+0.07005i$ & $10.06245-5.20522i$ & $-2.60644+2.33291i$ & $2$ & $-1.25055$ & $3$ \\
$-0.37881+0.02198i$ & $-0.07881+0.02198i$ & $-0.08454-0.09207i$ & $-0.02749+0.11816i$ & $0.06967-0.07005i$ & $10.06245+5.20522i$ & $2.60644+2.33291i$ & $3$ & $-1.25055$ & $3$ \\
$0.25000-0.04330i$ & $0.55000-0.04330i$ & $-0.77352-0.43943i$ & $-0.42648+0.56057i$ & $-0.10000-0.03454i$ & $-0.60291+1.15659i$ & $0.09819-0.00000i$ & $2$ & $-1.25055$ & $3$ \\
$0.25000+0.04330i$ & $0.55000+0.04330i$ & $-0.77352+0.43943i$ & $0.57352-0.56057i$ & $-1.10000+0.03454i$ & $-0.60291-1.15659i$ & $-0.09819-0.00000i$ & $3$ & $-1.25055$ & $3$ \\
$-0.45598-0.05509i$ & $-0.15598-0.05509i$ & $-0.87080-0.05853i$ & $1.09356+0.20136i$ & $-0.11080-0.03264i$ & $-1.36773+1.60135i$ & $0.51434+1.26882i$ & $2$ & $-0.86239$ & $4$ \\
$-0.45598+0.05509i$ & $-0.15598+0.05509i$ & $-0.87080+0.05853i$ & $1.09356-0.20136i$ & $-0.11080+0.03264i$ & $-1.36773-1.60135i$ & $-0.51434+1.26882i$ & $3$ & $-0.86239$ & $4$ \\
$0.25000+0.00965i$ & $0.55000+0.00965i$ & $-2.10000-0.03122i$ & $0.40000-0.16249i$ & $0.40000+0.17440i$ & $-0.08524-0.00000i$ & $-0.03396-0.00000i$ & $2$ & $-0.86239$ & $4$ \\
$0.25000-0.00965i$ & $0.55000-0.00965i$ & $-2.10000+0.03122i$ & $0.40000+0.16249i$ & $0.40000-0.17440i$ & $-0.08524+0.00000i$ & $0.03396-0.00000i$ & $3$ & $-0.86239$ & $4$ \\
$-0.75000+0.11275i$ & $-0.45000+0.11275i$ & $-1.90029-0.08059i$ & $-0.29971-0.08059i$ & $2.90000-0.06432i$ & $0.23171-0.00000i$ & $-0.58316+0.00000i$ & $2$ & $0.70428$ & $5$ \\
$-0.75000-0.11275i$ & $-0.45000-0.11275i$ & $-1.90029+0.08059i$ & $-0.29971+0.08059i$ & $2.90000+0.06432i$ & $0.23171+0.00000i$ & $0.58316+0.00000i$ & $3$ & $0.70428$ & $5$ \\
$-1.55828+0.03094i$ & $-1.25828+0.03094i$ & $-0.20115-0.04899i$ & $1.51859+0.03613i$ & $0.99912-0.04902i$ & $-0.07119+0.01370i$ & $-0.16566-0.04516i$ & $2$ & $0.70428$ & $5$ \\
$-1.55828-0.03094i$ & $-1.25828-0.03094i$ & $-0.20115+0.04899i$ & $1.51859-0.03613i$ & $0.99912+0.04902i$ & $-0.07119-0.01370i$ & $0.16566-0.04516i$ & $3$ & $0.70428$ & $5$ \\
$0.05720+0.13634i$ & $0.35720+0.13634i$ & $-2.06490-0.15055i$ & $1.37567+0.08642i$ & $-0.22517-0.20856i$ & $-0.23061+0.32198i$ & $-0.81379-0.24504i$ & $1$ & $1.02350$ & $6$ \\
$0.05720-0.13634i$ & $0.35720-0.13634i$ & $-2.06490+0.15055i$ & $1.37567-0.08642i$ & $-0.22517+0.20856i$ & $-0.23061-0.32198i$ & $0.81379-0.24504i$ & $4$ & $1.02350$ & $6$ \\
$0.25000-0.11861i$ & $0.55000-0.11861i$ & $-1.10000-0.14588i$ & $-0.42425+0.69155i$ & $0.22425-0.30845i$ & $-0.38487+0.76419i$ & $0.23725-0.00000i$ & $1$ & $1.02350$ & $6$ \\
$0.25000+0.11861i$ & $0.55000+0.11861i$ & $-1.10000+0.14588i$ & $-0.42425-0.69155i$ & $0.22425+0.30845i$ & $-0.38487-0.76419i$ & $-0.23725-0.00000i$ & $4$ & $1.02350$ & $6$ \\
$0.90991-0.30523i$ & $1.20991-0.30523i$ & $-3.10036-0.12527i$ & $0.11226+0.64836i$ & $0.36829+0.08736i$ & $0.51259+0.05215i$ & $7.25665+0.11304i$ & $1$ & $1.08128$ & $7$ \\
$0.90991+0.30523i$ & $1.20991+0.30523i$ & $-3.10036+0.12527i$ & $0.11226-0.64836i$ & $0.36829-0.08736i$ & $0.51259-0.05215i$ & $-7.25665+0.11304i$ & $4$ & $1.08128$ & $7$ \\
$0.25000+0.30649i$ & $0.55000+0.30649i$ & $-1.10000-0.12487i$ & $-0.60000-0.55900i$ & $0.40000+0.07089i$ & $0.25098-0.00000i$ & $5.34851+0.00000i$ & $1$ & $1.08128$ & $7$ \\
$0.25000-0.30649i$ & $0.55000-0.30649i$ & $-1.10000+0.12487i$ & $-0.60000+0.55900i$ & $0.40000-0.07089i$ & $0.25098-0.00000i$ & $-5.34851-0.00000i$ & $4$ & $1.08128$ & $7$ \\
$0.04768-0.36108i$ & $0.34768-0.36108i$ & $-0.70037+0.60978i$ & $0.07267-0.43720i$ & $-0.26765+0.54957i$ & $-0.45482+0.61880i$ & $0.94953+0.00473i$ & $0$ & $2.00622$ & $8$ \\
$-0.75000+0.35671i$ & $-0.45000+0.35671i$ & $-1.46212-0.58320i$ & $-0.10000-0.54702i$ & $2.26212+0.41680i$ & $-0.43735-0.51456i$ & $-0.93370-0.00000i$ & $0$ & $2.00622$ & $8$ \\
$0.25000+0.31675i$ & $0.55000+0.31675i$ & $-1.60000+0.08844i$ & $-0.10000-0.43539i$ & $0.40000-0.28655i$ & $0.42608-0.00000i$ & $-1.21388+0.00000i$ & $0$ & $2.35931$ & $9$ \\
$-1.08828-0.00000i$ & $-0.78828-0.00000i$ & $0.11378-0.00000i$ & $0.13139+0.44094i$ & $1.13139-0.44094i$ & $-1.85262+0.00000i$ & $-0.00000+0.50365i$ & $0$ & $2.35931$ & $9$ \\
$0.39331-0.09117i$ & $0.69331-0.09117i$ & $0.46344-0.00267i$ & $0.48984-0.16274i$ & $-2.53991+0.34774i$ & $-0.01886-0.06929i$ & $0.30512-0.18507i$ & $0$ & $2.69100$ & $10$ \\
$-1.36802+0.13801i$ & $-1.06802+0.13801i$ & $-0.26550+0.14338i$ & $1.63970-0.35821i$ & $0.56183-0.06119i$ & $-0.14181+0.59662i$ & $-0.50305-0.73047i$ & $0$ & $2.69100$ & $10$ \\
\hline\hline
\end{tabular}   } \end{center} \end{table}
\end{landscape}

\subsection{For a generic $\eta$}
Let us consider the $c=0$ solutions of
(\ref{BAE-Main-1})-(\ref{BAE-Main-5}). In this case, the
corresponding inhomogeneous $T-Q$ relation (\ref{T-Q-Main}) is
reduced to Baxter's homogeneous form \cite{bax2}. Obviously,
(\ref{BAE-Main-2}) is not necessary since $c=0$.

It follows from (\ref{BAE-Main-3}) and (\ref{BAE-Main-4}) that  for
$c=0$, the parameters $\{\mu_j\}$ and $\{\nu_j\}$ have to form the
pairs with either $\mu_j=\nu_k$ or $\mu_j=\nu_k-\eta$. Suppose \bea
\mu_j&=&\nu_j\stackrel{{\rm Redef}}{=}\l_{M_1+j},\quad j=1,\ldots,\bar m,\quad {\rm and}\quad 0\leq \bar m\leq M,\no\\
\mu_{\bar m+k}&=&\nu_{k+\bar m}-\eta,\quad k=1,\ldots,M-\bar m.\label{Pairs-1} \eea
Combining (\ref{Pairs-1}) with (\ref{BAE-Main-1}), we have \bea
(\frac{N}2-\bar m-M_1)\eta=l_1\tau+m_1.\label{P-Con} \eea
\begin{itemize}
\item {\bf Even $N$ case}. Suppose $N=2\bar M$. Because $\tau$ and $\eta$ are generic complex numbers, the only solution
to (\ref{P-Con}) is \bea l_1=m_1=0, \quad N=2\bar M=2(M_1+\bar m). \eea
The resulting  $T-Q$ relation (\ref{T-Q-Main}) is reduced to
Baxter's one
\begin{eqnarray}
\Lambda(u)&=&e^{i\phi}\frac{\sigma^N(u+\eta)}{\sigma^N(\eta)}\frac{Q(u-\eta)}{Q(u)}
+e^{-i\phi}\frac{\sigma^N(u)}{\sigma^N(\eta)}\frac{Q(u+\eta)}{Q(u)},\label{T-Q-Baxter-1}\\
Q(u)&=&\prod_{l=1}^{M_1}\frac{\s(u-\l_l)}{\s(\eta)}\prod_{k=1}^{\bar m}\frac{\s(u-\nu_k)}{\s(\eta)}=
\prod_{l=1}^{\bar M}\frac{\s(u-\l_l)}{\s(\eta)}.\label{Q-3}
\end{eqnarray}
The resulting BAEs and the selection rule thus read
\begin{eqnarray}
\frac{\sigma^N(\lambda_j+\eta)}{\sigma^N(\lambda_j)}&=&-e^{-2i\phi}\frac{{Q}(\lambda_j+\eta)}
{{ Q}(\lambda_j-\eta)},\quad j=1,\ldots,\bar M, \label{Generic-BAE-1}\\
e^{i\phi}\,\prod_{j=1}^{\bar M}\frac{\sigma(\lambda_j+\eta)}{\sigma(\lambda_j)}&=&e^{\frac{2i\pi k}{N}},
\quad k=1,\ldots,N.\label{Generic-BAE-2}
\end{eqnarray}
Some remarks are in order. The BAEs (\ref{Generic-BAE-1}) are just
those  obtained in Refs. \cite{bax2,xyz2}, while the relation
(\ref{Generic-BAE-2}) gives rise to that the parameter $\phi$ takes
a discrete value labeled by $k=1,\ldots,N$.  On the other hand,
$c\neq0$ and $\mu_j\neq\nu_k$ for arbitrary $j,k$ may not lead to
new solutions but different parameterizations as discussed by Baxter
\cite{bax04} that $\bar M=N/2$ already gives a complete set of
solutions for even $N$. To show this clearly, the numerical
solutions for $N=4$ and random choice of $\eta$ and $\tau$ are listed
in Table 3.
\begin{table}
\caption{\label{5N4} Numerical solutions of the BAEs
(\ref{Generic-BAE-1})-(\ref{Generic-BAE-2}) for $N=4$, $\eta=0.4$,
$\tau=i$. The eigenvalues $E_n$ are exactly the same to those from
the exact diagonalization. $n$ denotes the number of the energy
levels.}
\begin{center}{\scriptsize
\begin{tabular}{ cc|c|c|c|c} \hline \hline
$\lambda_1$ & $\lambda_2$  & $\phi$ & $k$ & $E_n$ & $n$ \\
\hline
$0.80000+0.11349i$ &$0.80000+0.88651i$ & $2.51327$   &$2$ & $-3.21353$ & $1$ \\
$0.80000+0.00000i$ &$0.80000+0.50000i$ & $1.25664$   &$1$ & $-2.34227$ & $2$ \\
$0.80000+0.00000i$ &$0.30000+0.50000i$  &  $1.25664$  &$1$ & $-1.71217$ & $3$ \\
$0.30000+0.00000i$ &$0.80000+0.00000i$ &$0$ &$0$ & $-0.61387$ & $4$ \\
$0.30000+0.70000i$ &$0.80000+0.80000i$ & $3.76991$   &$3$ & $0.00000$ & $5$ \\
$0.30000+0.30000i$ &$0.80000+0.20000i$ & $1.25664$  &$1$ & $0.00000$ & $5$ \\
$0.30000+0.86676i$ &$0.80000+0.13324i$ & $2.51327$   &$2$ & $0.00000$ & $5$ \\
$0.30000+0.13324i$ &$0.80000+0.86676i$  & $2.51327$   &$2$ & $0.00000$ & $5$ \\
$0.62340+0.25000i$ &$0.97660+0.25000i$  &  $1.25664$  &$1$ & $0.00000$ & $5$ \\
$0.62340+0.75000i$ &$0.97660+0.75000i$  &  $3.76991$  &$3$ & $0.00000$ & $5$ \\
$0.6$ &$1.0$  & $0$ &$0$ & $0.00000$ & $5$ \\
$0.03367+0.50000i$ &$0.56633+0.50000i$  &$2.51327$  &$2$ & $0.58230$ & $6$ \\
$0.30000+0.50000i$ &$0.80000+0.50000i$  &$2.51327$  &$2$ & $0.61387$ & $7$ \\
$0.30000+0.00000i$ &$0.80000+0.50000i$  & $1.25664$ &$1$ & $1.71217$ & $8$ \\
$0.30000+0.00000i$ &$0.30000+0.50000i$  & $1.25664$ &$1$ & $2.34227$ & $9$ \\
$0.30000+0.16022i$ &$0.30000+0.83978i$  &$2.51327$  &$2$ & $2.63122$ & $10$ \\
\hline\hline \end{tabular} } \end{center}\end{table}\\

\item {\bf Odd $N$  case}. Since $\tau$ and $\eta$ are generic complex numbers,
(\ref{P-Con}) cannot be satisfied for any odd $N$. This means that
the $c=0$ solution of the BAEs (\ref{BAE-Main-1})-(\ref{BAE-Main-5})
does not exist for an odd $N$ and generic $\tau$ and $\eta$.
\end{itemize}

\subsection{For some degenerate values of $\eta$}

For some degenerate values of $\eta$, the $c=0$ solutions indeed
exist no matter $N$ is even or odd. In this case, (\ref{BAE-Main-2}) 
is not necessary, and the parameters
$\eta$ and $\tau$ are no longer independent but have to obey the
relation (\ref{P-Con}). This implies that if the crossing parameter
$\eta$ takes some discrete values \bea \eta=\frac{2l_1}{N-2\bar
M}\tau+\frac{2m_1}{N-2\bar M}, \label{Discrete} \eea for any
non-negative integer $\bar M=M_1+\bar m$ and any integers $l_1$ and
$m_1$, our generalized $T-Q$ relation (\ref{T-Q-Main}) is reduced to the
conventional one \cite{bax2,xyz2}
\begin{eqnarray}
\Lambda(u)&=&e^{2i\pi l_1u+i\phi}\frac{\sigma^N(u+\eta)}{\sigma^N(\eta)}\frac{Q(u-\eta)}{Q(u)}
+e^{-2i\pi l_1(u+\eta)-i\phi}\frac{\sigma^N(u)}{\sigma^N(\eta)}\frac{Q(u+\eta)}{Q(u)},\label{T-Q-Baxter}
\end{eqnarray} where the $Q$-function is given by (\ref{Q-3}). The $\bar M+1$ parameters $\phi$ and $\{\lambda_j\}$  satisfy the associated BAEs
\bea
 &&e^{\{2i\pi(2l_1\lambda_j+l_1\eta)+2i\phi\}}
 \frac{\sigma^N(\lambda_j+\eta)}{\sigma^N(\lambda_j)}=-\frac{Q(\lambda_j+\eta)}{Q(\lambda_j-\eta)},\quad j=1,\ldots,\bar{M},
 \label{Special-BAE-1}\\[6pt]
&&e^{i\phi}\,\prod_{j=1}^{\bar M}\frac{\sigma(\lambda_j+\eta)}{\sigma(\lambda_j)}=e^{\frac{2i\pi k}{N}},\quad k=1,\ldots,N.\label{Special-BAE-2}
\eea

%%%%%%%%%%%%%%%%%%%%%%%%%%%%%%%%%%%%%%%%%%%%%%%%%%%%%%%%%%%%%%%
%                                                             %
%  4. Results for the XYZ chain with antiperiodic  boundary   %
%                       condition                             %
%                                                             %
%                                                             %
%                                                             %
%%%%%%%%%%%%%%%%%%%%%%%%%%%%%%%%%%%%%%%%%%%%%%%%%%%%%%%%%%%%%%%

\section{Results for the XYZ chain with anti-periodic  boundary condition }
\label{APBC}
\setcounter{equation}{0}

\subsection{Functional relations}
Now let us turn to the XYZ spin chain described by the Hamiltonian
(\ref{xyzh}) but with the anti-periodic boundary condition
(\ref{Anti-periodic}). Its integrability is associated with the
mutually commutative transfer matrix $t^{(a)}(u)$ given by \bea
 t^{(a)}(u)=tr_0\{\sigma^x_0\,T_0(u)\}.\label{anti-Transfer}
\eea Following the method introduced in Section 2, we can derive the
following functional relations
\bea
 t^{(a)}(\theta_j)t^{(a)}(\theta_j-\eta)&=&-a(\theta_j)\,d(\theta_j-\eta),\quad j=1,\ldots N,\label{Identity-Anti-1}\\[6pt]
 \prod_{j=1}^N t^{(a)}(\theta_j)&=&\lt\{\prod_{j=1}^Na(\theta_j)\rt\} \times U^x,\label{Identity-Anti-2}\\[6pt]
 t^{(a)}(u+1)&=&(-1)^{N-1}\,t^{(a)}(u),\label{Identity-Anti-3}\\[6pt]
 t^{(a)}(u+\tau)&=&(-1)^Ne^{-2i\pi\{Nu+N\frac{\eta+\tau}{2}-\sum_{l=1}^N\theta_l\}}\,t^{(a)}(u),\label{Identity-Anti-4}
\eea where the operator $U^x$ is given by (\ref{Invarinat-2}). 
It is easy to check that \bea [t^{(a)}(u),\,U^x]=0,\quad
(U^x)^2={\rm id},\no \eea which implies that the eigenvalue of the
operator $U^x$ takes the values $\pm 1$ and can be diagonalized with
the transfer matrix $t^{(a)}(u)$ simultaneously.  Let us denote the
eigenvalue of the transfer matrix $t^{(a)}(u)$ as $\Lambda(u)$.
(\ref{Identity-Anti-1})-(\ref{Identity-Anti-4}) enable us to derive
the following functional relations \bea
 \Lambda(\theta_j)\Lambda(\theta_j-\eta)&=&-a(\theta_j)\,d(\theta_j-\eta),\quad j=1,\ldots N,\label{Eigen-Identity-Anti-1}\\[6pt]
 \prod_{j=1}^N \Lambda(\theta_j)&=&\pm \prod_{j=1}^Na(\theta_j),\label{Eigen-Identity-Anti-2}\\[6pt]
 \Lambda(u+1)&=&(-1)^{N-1}\Lambda(u),\label{Eigen-Identity-Anti-3}\\[6pt]
 \Lambda(u+\tau)&=&(-1)^Ne^{-2i\pi\{Nu+N\frac{\eta+\tau}{2}-\sum_{l=1}^N\theta_l\}}\Lambda(u).\label{Eigen-Identity-Anti-4}
\eea   The analyticity of the $R$-matrix  implies the following
analytic property of $\Lambda(u)$
\begin{eqnarray}
 \Lambda(u) \mbox{ is an entire
function of $u$}.\label{Eigen-Anti-1}
\end{eqnarray}

\subsection{$T-Q$ relation}

As for the periodic case,
(\ref{Eigen-Identity-Anti-1})-(\ref{Eigen-Anti-1}) allow us to
determine the eigenvalues of the transfer matrix $t^{(a)}(u)$. After
taking the homogeneous limit $\theta_j\rightarrow 0$, we obtain the
following inhomogeneous $T-Q$ relation
\begin{eqnarray}
\Lambda(u)&=&e^{\{i\pi(2l_1+1)u+i\phi\}}\frac{\sigma^N(u+\eta)}{\sigma^N(\eta)}\frac{Q_1(u-\eta)Q(u-\eta)}{Q_2(u)Q(u)}\no\\[6pt]
&&-\frac{e^{-i\pi(2l_1+1)(u+\eta)-i\phi}\sigma^N(u)}{\sigma^N(\eta)}\frac{Q_2(u+\eta)Q(u+\eta)}{Q_1(u)Q(u)}\nonumber\\[6pt]
&&+\frac{c\,e^{i\pi u}\sigma^N(u+\eta)\sigma^N(u)}{Q_1(u)Q_2(u)Q(u)\sigma^N(\eta)\sigma^N(\eta)},\label{T-Q-Main-Anti}
\end{eqnarray}
where $l_1$ is a certain integer, the $Q$-functions $Q_1(u)$, $Q_2(u)$ and $Q(u)$ are given by (\ref{Q-1})-(\ref{Q-2}).
The $N+2$ parameters $c$, $\phi$, $\{\mu_j|j=1,\ldots,M\}$, $\{\nu_j|j=1,\ldots,M\}$ and $\{\l_j|j=1,\ldots,M_1\}$ satisfy the associated BAEs
\begin{eqnarray}
&&\hspace{-1.2truecm}(\frac{N}2-M-M_1)\eta-\sum_{j=1}^M(\mu_j-\nu_j)=(l_1+\frac{1}{2})\tau+m_1, \quad l_1,\,m_1\in Z, \label{BAE-Anti-1}\\[6pt]
&&\hspace{-1.2truecm}\frac{N}{2}\eta
+\sum_{j=1}^M(\mu_j+\nu_j)+\sum_{j=1}^{M_1}\l_j=\frac{1}{2}\tau+m_2,\quad m_2\in Z,\label{BAE-Anti-2}\\[6pt]
&&\hspace{-1.2truecm}\frac{c\,e^{\{2i\pi(l_1+1)\mu_j+2i\pi(l_1+\frac{1}{2})\eta+i\phi\}}
\sigma^N(\mu_j\hspace{-0.02truecm}+\hspace{-0.02truecm}\eta)}{\sigma^N(\eta)}=
Q_2(\mu_j)Q_2(\mu_j\hspace{-0.02truecm}+\hspace{-0.02truecm}\eta)
Q(\mu_j\hspace{-0.02truecm}+\hspace{-0.02truecm}\eta),\,j=1,\ldots,M,\label{BAE-Anti-3}\\[6pt]
&&\hspace{-1.2truecm}\frac{c\,e^{\{-2i\pi l_1\nu_j-i\phi\}}\sigma^N(\nu_j)}{\sigma^N(\eta)}=
-Q_1(\nu_j)Q_1(\nu_j-\eta)Q(\nu_j-\eta),\,j=1,\ldots,M,\label{BAE-Anti-4}\\[6pt]
&&\hspace{-1.2truecm}e^{i\pi(2l_1+1)(2\l_j+\eta)+2i\phi}\frac{\s^N(\l_j+\eta)}{\s^N(\l_j)}-\frac{Q_2(\l_j)Q_2(\l_j+\eta)Q(\l_j+\eta)}
{Q_1(\l_j)Q_1(\l_j-\eta)Q(\l_j-\eta)}\no\\[6pt]
&&\qquad\qquad =\frac{-c\,e^{2i\pi(l_1+1)\l_j+i\pi(2l_1+1)\eta+i\phi}\s^N(\l_j+\eta)}
{Q_1(\l_j)Q_1(\l_j-\eta)Q(\l_j-\eta)\s^N(\eta)},\,j=1\ldots,M_1,
\label{BAE-Anti-5}
\end{eqnarray}
and the  selection rule

\bea
\Lambda(0)=e^{i\phi}\lt\{\prod_{j=1}^M\frac{\sigma(\mu_j+\eta)}{\sigma(\nu_j)}\rt\}
\lt\{\prod_{j=1}^{M_1}\frac{\sigma(\l_j+\eta)}{\sigma(\l_j)}\rt\}=e^{\frac{i\pi
k}{N}},\quad k=1,\ldots,2N.\label{Selection-Anti} \eea \noindent The
eigenvalue of the Hamiltonian (\ref{xyzh}) with the anti-periodic
boundary condition is then given by
\begin{eqnarray}
E&=&\frac{\sigma(\eta)}{\sigma'(0)}\lt\{\sum_{j=1}^M[\zeta(\nu_j)-\zeta(\mu_j+\eta)]+
\sum_{j=1}^{M_1}[\zeta(\l_j)-\zeta(\l_j+\eta)]\rt.\no\\[6pt]
&&\quad\quad\quad\quad+\lt.\frac{1}{2}N\zeta(\eta)+2i\pi
(l_1+\frac{1}{2})\rt\}.
\end{eqnarray}

For a generic $\eta$, in contrast to the periodic case, there does
not exist the $c=0$ solution of the BAEs
(\ref{BAE-Anti-1})-(\ref{BAE-Anti-5}) no matter $N$ is  even or odd.
However, when $\eta$ takes some discrete values labeled by a
non-negative integer $\bar M$ and two integers $l_1$ and $m_1$
\bea
 \eta=\frac{2l_1+1}{N-2\bar{M}}\tau +\frac{2m_1}{N-2\bar{M}},\quad l_1,m_1\in Z,\label{M-Anti-2}
\eea the $c=0$ solutions of the BAEs
(\ref{BAE-Anti-1})-(\ref{BAE-Anti-5}) do exist. In this case, the $T-Q$
relation (\ref{T-Q-Main-Anti}) is  reduced to the conventional one
\begin{eqnarray}
\Lambda(u)\hspace{-0.06truecm}&=&\hspace{-0.06truecm}e^{2i\pi (l_1+\frac{1}{2})u+i\phi}
\frac{\sigma^N(u\hspace{-0.06truecm}+\hspace{-0.06truecm}\eta)}{\sigma^N(\eta)}
\frac{Q(u\hspace{-0.06truecm}-\hspace{-0.06truecm}\eta)}{Q(u)}
\hspace{-0.06truecm}-\hspace{-0.06truecm}e^{-2i\pi (l_1+\frac{1}{2})(u+\eta)-i\phi}\frac{\sigma^N(u)}{\sigma^N(\eta)}
\frac{Q(u\hspace{-0.06truecm}+\hspace{-0.06truecm}\eta)}{Q(u)},\label{T-Q-Baxter-Anti}
\end{eqnarray} with the associated BAEs and selection rule
\bea
 &&e^{\{2i\pi((2l_1+1)\lambda_j+(l_1+\frac{1}{2})\eta)+2i\phi\}}
 \frac{\sigma^N(\lambda_j+\eta)}{\sigma^N(\lambda_j)}=\frac{Q(\lambda_j+\eta)}{Q(\lambda_j-\eta)},\quad j=1,\ldots,\bar M, \label{Special-BAE-Anti-1}\\[6pt]
&&e^{i\phi}\,\prod_{j=1}^{\bar M}\frac{\sigma(\lambda_j+\eta)}{\sigma(\lambda_j)}=e^{\frac{i\pi k}{N}},\quad k=1,\ldots,2N.\label{Special-BAE-Anti-2}
\eea

%%%%%%%%%%%%%%%%%%%%%%%%%%%%%%%%%%%%%%%%%%%%%%%%%%%%%%%%%%%%%%%
%                                                             %
%  5. Conclusions                                             %
%                                                             %
%                                                             %
%                                                             %
%%%%%%%%%%%%%%%%%%%%%%%%%%%%%%%%%%%%%%%%%%%%%%%%%%%%%%%%%%%%%%%

\section{Conclusions}
\label{Con}

The spin-${\frac 12}$ XYZ model  described by the Hamiltonian
(\ref{xyzh}) with the periodic boundary condition
(\ref{Periodic-BC}) and the anti-periodic boundary condition
(\ref{Anti-periodic}) are studied via the off-diagonal Bethe ansatz
method \cite{cao1,cao2,cao2-1}. The eigenvalues of the transfer
matrices are given in terms of the inhomogeneous $T-Q$ relations
(\ref{T-Q-Main}) and (\ref{T-Q-Main-Anti}) which allow us to treat
both even $N$ and odd $N$ cases in an unified framework. For a
generic crossing parameter $\eta$, our solution can be reduced to
Baxter's solution only for the periodic chain and even $N$, while
for all the other cases (the periodic chain with odd $N$ and the
anti-periodic chain), an extra
inhomogeneous term (the third term in (\ref{T-Q-Main}) or
(\ref{T-Q-Main-Anti})) has to be included in the $T-Q$ relations.
However, if the crossing parameter $\eta$ takes some degenerate
values ((\ref{Discrete}) for the periodic case and (\ref{M-Anti-2})
for the antiperiodic case), the corresponding $T-Q$ relation
can be reduced to the conventional one. It should be emphasized that
these degenerate points become dense in the whole complex
$\eta$-plane in the thermodynamic limit ($N\rightarrow \infty$).
This enables one to obtain the thermodynamic properties (up to the
order of $O(N^{-2})$) \cite{Li14} for generic values of $\eta$ via
the conventional thermodynamic Bethe ansatz methods \cite{yb,Tak99}.

%%%%%%%%%%%%%%%%%%%%%%%%%%%%%%%%%%%%%%%%%%%%%%%%%%%%%%%%%%%%%%%
%                                                             %
%  Acknowledgments                                            %
%                                                             %
%%%%%%%%%%%%%%%%%%%%%%%%%%%%%%%%%%%%%%%%%%%%%%%%%%%%%%%%%%%%%%%
\section*{Acknowledgments}

The financial supports from  the National Natural Science Foundation
of China (Grant Nos. 11174335, 11031005, 11375141, 11374334), the
National Program for Basic Research of MOST (973 project under grant
No. 2011CB921700), the State Education Ministry of China (Grant No.
20116101110017) and BCMIIS are gratefully acknowledged. Two of the
authors (W.\,-L Yang and K. Shi) would like to thank IoP/CAS for the
hospitality and they enjoyed during their visit there.

\section*{Appendix A: Identities of the elliptic functions }
\setcounter{equation}{0}
\renewcommand{\theequation}{A.\arabic{equation}}
The following identities for the elliptic functions defined by
(\ref{Function-a-b})-(\ref{Function}) are  quite useful in the
derivations \bea
&&\s(u+x)\s(u-x)\s(v+y)\s(v-y)-\s(u+y)\s(u-y)\s(v+x)\s(v-x)\no\\[6pt]
&&~~~~~~=\s(u+v)\s(u-v)\s(x+y)\s(x-y),\label{Identity-fuction}\\[6pt]
&&\s(2u)=\frac{2\s(u)\s(u+\frac{1}{2})\s(u+\frac{\tau}{2})
\s(u-\frac{1}{2}-\frac{\tau}{2})}{\s(\frac{1}{2})\s(\frac{\tau}{2})
\s(-\frac{1}{2}-\frac{\tau}{2})},\label{Identity-f-1}\\[6pt]
&&\frac{\s(u)}{\s(\frac{\tau}{2})}=\frac{ \theta\lt[\begin{array}{l}
0\\\frac{1}{2}\end{array}\rt](u,2\tau)\,\,\theta\lt[\begin{array}{l}
\frac{1}{2}\\[2pt]\frac{1}{2}\end{array}\rt](u,2\tau)}
{\theta\lt[\begin{array}{l}
0\\\frac{1}{2}\end{array}\rt](\frac{\tau}{2},2\tau)\,\,
\theta\lt[\begin{array}{l}
\frac{1}{2}\\[2pt]\frac{1}{2}\end{array}\rt]
(\frac{\tau}{2},2\tau)},\label{Identity-f-2} \\[6pt]
&&\theta
\lt[\begin{array}{c}\frac{1}{2}\\[2pt]\frac{1}{2}\end{array}\rt]
(2u,2\tau)=\theta
\lt[\begin{array}{c}\frac{1}{2}\\[2pt]\frac{1}{2}\end{array}\rt]
(\tau,2\tau)\,\times\,\frac{\s(u)\s(u+\frac{1}{2})}
{\s(\frac{\tau}{2})\s(\frac{1}{2}+\frac{\tau}{2})},\label{Identity-f-3}\\[6pt]
&&\theta \lt[\begin{array}{c}0\\\frac{1}{2}\end{array}\rt]
(2u,2\tau)=\theta \lt[\begin{array}{c}0\\\frac{1}{2}\end{array}\rt]
(0,2\tau)\,\times\,\frac{\s(u-\frac{\tau}{2})\s(u+\frac{1}{2}+\frac{\tau}{2})}
{\s(-\frac{\tau}{2})\s(\frac{1}{2}+\frac{\tau}{2})}.\label{Identity-f-4}
\eea

\section*{Appendix B: Trigonometric limit}
\setcounter{equation}{0}
\renewcommand{\theequation}{B.\arabic{equation}}
The results of the XXZ spin chain can be recovered by taking the
limit $\tau\rightarrow +i\infty$ of the XYZ model. Here we take the
periodic case as an example. Its generalization to the anti-periodic
case is straightforward.

The definition of the elliptic functions
(\ref{Function-a-b})-(\ref{Function}) implies
\bea \s(u+\frac{\tau}{2})=e^{-i\pi(u+\frac{1}{2}+\frac{\tau}{4})}
\,\,\theta\lt[\begin{array}{c}0\\\frac{1}{2}\end{array}\rt](u,\tau),
\label{AD-1}
\eea
and the following asymptotic behaviors \bea
&&\lim_{\tau\rightarrow
+i\infty}\s(u)=-2e^{\frac{i\pi\tau}{4}}\sin \pi
u+\ldots,\label{AD-2}\\[6pt]
&&\lim_{\tau\rightarrow
+i\infty}\theta\lt[\begin{array}{c}0\\\frac{1}{2}\end{array}\rt](u,\tau)=
1+\ldots.\label{AD-3}\eea The above asymptotic behaviors lead to the
well-known XXZ $R$-matrix
\bea \lim_{\tau\rightarrow
+i\infty}R(u)=\frac{1}{\sin\,\pi\eta}\lt(\begin{array}{cccc}\sin\pi(u+\eta)
&&&\\&\sin\,\pi u&\sin\,\pi\eta&\\
&\sin\,\pi\eta&\sin\,\pi u&\\&&&\sin\pi(u+\eta)\end{array}\rt).
\label{t-r-matrix}\eea
The resulting $R$-matrix gives rise to the associated
asymptotic behaviors of the resulting transfer matrix, which are the
counterparts of the quasi-periodic properties
(\ref{quasi-transfer-1}) and (\ref{quasi-transfer-2}), \bea
&&\hspace{-1.2truecm}t(u+1)=(-1)^Nt(u),\label{Trig-Per}\\[6pt]
&&\hspace{-1.2truecm}t(u)\stackrel{u\rightarrow -i\infty}{=} \frac{e^{i\pi(Nu-\sum_{l=1}^N\theta_l+\frac{N}{2}\eta)}}{(2\sin\pi\eta)^N}
\lt(e^{\frac{i\pi\eta}{2}\sum_{l=1}^N\s^z_l}+e^{\frac{-i\pi\eta}{2}\sum_{l=1}^N\s^z_l}\rt)+\ldots,\\[6pt]
&&\hspace{-1.2truecm}t(u)\stackrel{u\rightarrow +i\infty}{=}(-1)^N \frac{e^{i\pi(-Nu+\sum_{l=1}^N\theta_l-\frac{N}{2}\eta)}}{(2\sin\pi\eta)^N}
\lt(e^{\frac{i\pi\eta}{2}\sum_{l=1}^N\s^z_l}+e^{\frac{-i\pi\eta}{2}\sum_{l=1}^N\s^z_l}\rt)+\ldots.
\eea
Since the total spin operator
$S^z=\frac{1}{2}\sum_{l=1}^N\s^z_l$ commutes with the transfer matrix
in the trigonometric limit, one can decompose the whole Hilbert
space into subspaces according to the eigenvalues of $S^z$ \bea
\Cb^2\otimes
\Cb^2\otimes\cdots\Cb^2=\bigoplus_{i=0}^N{\cal{H}}^{(i)},\quad
S^z{\cal{H}}^{(i)}=(\frac{N}{2}-i){\cal{H}}^{(i)}. \eea The
eigenvalue $\Lambda(u)$ in the subspace ${\cal{H}}^{(M)}$ has the
following asymptotic behaviors \bea
&&\Lambda(u+1)=(-1)^N\Lambda(u),\label{Trignom}\\[6pt]
&&\Lambda(u) \stackrel{u\rightarrow -i\infty}{=}\frac{e^{i\pi(Nu-\sum_{l=1}^N
     \theta_l)}}{(2\sin\pi\eta)^N}(e^{i\pi(N-M)\eta}+e^{i\pi M\eta})+\ldots,\label{Trignom-1}\\[6pt]
&&\Lambda(u) \stackrel{u\rightarrow +i\infty}{=}(-1)^N\frac{e^{i\pi(-Nu+\sum_{l=1}^N
     \theta_l)}}{(2\sin\pi\eta)^N}(e^{i\pi(-N+M)\eta}+e^{-i\pi M\eta})+\ldots.\label{Trignom-2}
\eea The limits of the identities (\ref{Eigen-identity}) become
\begin{eqnarray}
&&\Lambda(\theta_j)\Lambda(\theta_j-\eta) =\bar{a}(\theta_j)
\bar{d}(\theta_j-\eta),\,j=1,\ldots,N,\label{Eigen-identity-1}\\[6pt]
&&\bar{a}(u)=\prod_{l=1}^N\frac{\sin\pi(u-\theta_l+\eta)}{\sin\pi\eta},\quad \bar{d}(u)=\bar{a}(u-\eta)=\prod_{l=1}^N\frac{\sin\pi(u-\theta_l)}{\sin\pi\eta}.\no
\end{eqnarray} The solutions of (\ref{Eigen-Per-1}),  (\ref{Trignom})-(\ref{Eigen-identity-1}) in the subspace
${\cal{H}}^{(M)}$ (naturally  $c=0$) can be given by the usual $T-Q$
relation \bea
&&\Lambda(u)=\bar{a}(u)\frac{\bar{Q}(u-\eta)}{\bar{Q}(u)}+ \bar{d}(u)\frac{\bar{Q}(u+\eta)}{\bar{Q}(u)},\\[6pt]
&&\bar{Q}(u)=\prod_{l=1}^M\frac{\sin\pi(u-\lambda_l)}{\sin\pi\eta},\quad M=0,1,\ldots,N,
\eea where the Bethe roots $\{\lambda_l\}$ satisfy the conventional
Bethe ansatz equations \cite{yb}.

\end{document}